# Background charge fluctuation in a GaAs quantum dot device

S. W. Jung,[1,2] T. Fujisawa,[1,3] Y. H. Jeong[2] and Y. Hirayama[1,4]

[1] *NTT Basic Research Laboratories, NTT Corporation, 3-1 Morinosato-Wakamiya, Atsugi, 243-0198, Japan*
[2] *Pohang University of Science and Technology, Pohang, Kyungpook, 790-784, Korea*
[3] *Tokyo Institute of Technology, 2-12-1, O-okayama, Meguro, Tokyo, 152-8550, Japan*
[4] *SORST-JST, 4-1-8, Honmachi, Kawaguchi, Saitama, 331-0012, Japan*

We investigate background charge fluctuation in a GaAs quantum dot device by measuring 1/f noise in the single-electron tunneling current through the dot. The current noise is understood as fluctuations of the confinement potential and tunneling barriers. The estimated potential fluctuation increases almost linearly with temperature, which is consistent with a simple model of the 1/f noise. We find that the fluctuation increases very slightly when electrons are injected into excited states of the quantum dot.

A quantum dot (QD) in the Coulomb blockade (CB) regime allows us to study tunable quantum states.[1] Such a QD also behaves as a single-electron transistor which is widely used to detect a single electron, single spin, and a single photon.[2-4] A common problem in these devices is 1/f noise, which appears in a current through the device and degrades the stability and coherency of the system.[5,6] Although the microscopic origin of the noise is not known well, it is generally believed that charge distribution of electron traps in the device fluctuates randomly with time (background charge fluctuation).[7,8] An understanding of 1/f noise is an important step in improving device characteristics. Many studies of 1/f noise have been carried out on single-electron devices and semiconductor nano-structures.[9-12] However, most works have focused on the classical CB regime, where energy quantization is not important. In this work, we studied the 1/f noise in the nonlinear transport regime, and investigated the fluctu-ation of the electrostatic potential and tunneling rate of the device.

The QD was fabricated in an AlGaAs/GaAs heterostructure using a dry etching technique and Schottky fine gates [See Fig. 1(a)].[6] All measurements were performed in a dilution refrigerator in the temperature range of $T$ = 20 - 300 mK at zero magnetic field. We applied negative voltages, $V_L$ and $V_R$, to two gate electrodes to form a single dot (shown by a white circle). The QD shows clear CB characteristics with a charging energy of $E_C$ ~ 2 meV and clear quantized energy states with typical energy spacing of ~ 100 μeV. By adjusting $V_L$ and $V_R$, the tunneling rates for the left and right barriers, $\Gamma_L$ and $\Gamma_R$ respectively, are made strongly asymmetric ($\Gamma_L$ ~ $3\times10^8$/s << $\Gamma_R$ ~ $6\times10^{10}$/s, which are determined from the $V_L$ and $V_R$ dependence of the non-linear current).[7] Figure 1(b) shows a single-electron tunneling peak obtained when $V_L$ was swept to change the electrostatic potential. $\Gamma_L$ does not change so much in this small sweeping range. In the nonlinear transport regime, the current increases or decreases stepwise when a quantized state enters or leaves the transport window between the electrochemical potential of the electrodes, $\mu_s$ and $\mu_d$, respectively, for the source and drain.[1] The electrochemical potential of the QD, $\varepsilon_{i,j}$, is defined as $\varepsilon_{i,j}$ = $U(N+1, i) - U(N, j)$, where $U(N, j)$ is the total energy of the $j$-th state of $N$-electron QD. We use $i$ ($j$) = 0 for the ground state and $i$ ($j$) > 0 for excited states. The current steps indicated by solid lines in Fig. 1(b) are associated with the energy alignment of $\varepsilon_{i,j}$ to $\mu_s$ or $\mu_d$. Since we have chosen $\Gamma_L$ << $\Gamma_R$, only the resonances of excited states with the source potential ($\mu_s = \varepsilon_{i, j}$) through the thick barrier appear if the energy relaxation inside the dot is efficient.[13] For this specific peak, three excited states ($j$ = 1, 2, and 3) for the $N$-electron QD and one excited state ($i$ = 1) for the ($N$+1)-electron QD are observed in the transport energy window of 700 μeV. In this nonlinear transport regime, transport characteristics can be parameterized by the electrochemical potential $\varepsilon_{i,j}$ and tunneling rates $\Gamma_L$ and $\Gamma_R$. In what follows, we discuss how much these parameters fluctuate.

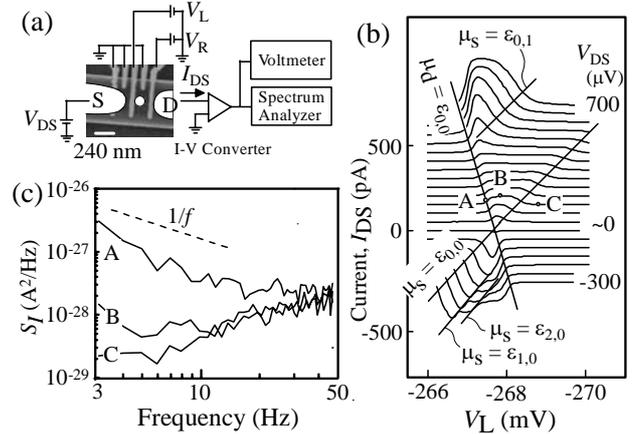

Fig. 1. (a) Schematic of the setup for current noise measurement. The source, drain, and quantum dot regions are schematically illustrated in the scanning electron micrograph. (b) Current profile of a single-electron tunneling peak when $V_L$ is swept. Each trace is offset for clarity. Lines show current steps associated with the energy alignment. (c) Frequency spectrum of the current fluctuation. The dashed line shows 1/f dependence.

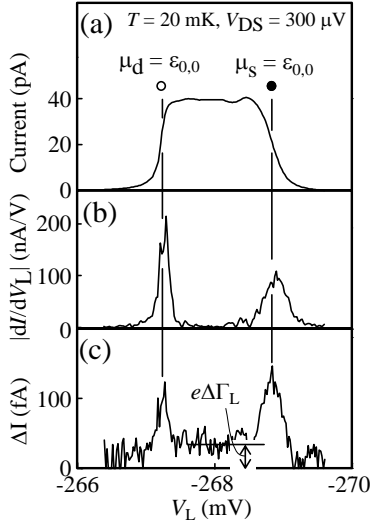

Fig. 2. (a) Single electron current profile obtained at $V_{DS} = 300$ μV. (b) The derivative of the current spectra. (c) The magnitude of the current fluctuation integrated from 5 to 45 Hz. Vertical lines indicate the conditions, $\mu_d = \varepsilon_{0,0}$ and $\mu_s = \varepsilon_{0,0}$.

We measured the fluctuation of the tunneling current with a spectrum analyzer. Figure 1(c) shows typical current noise spectra, $S_I(f)$, measured at different condition, labeled A, B, and C in Fig. 1(b). The spectrum at C was measured at zero current in the CB region, and indicates the noise of our measurement system. The spectrum at A shows a 1/f like spectrum at low frequency. However, the spectrum at B shows smaller noise even though the averaged current at B is higher than at A. We estimated the magnitude of current fluctuation $\Delta I$ by integrating the spectrum in a limited frequency range between 5 and 45 Hz, i.e.,

$$\Delta I = \sqrt{\int_5^{45} [S_I(f) - S_{I,CB}(f)] df} \cdot \qquad (1)$$

Here, the background noise spectrum, $S_{I,CB}(f)$, obtained in the CB regime, is subtracted to estimate the noise only from the sample. We assume that this excess current fluctuation is dominated by the background charge fluctuation, which typically shows 1/f frequency spectrum. Shot noise is known as an intrinsic noise for tunneling devices, but the expected white noise spectrum of about $1.2 \times 10^{-29}$ A$^2$/Hz at 40 pA is smaller than the observed spectrum. The noise in the control voltages may not be eliminated completely, but it is irrelevant to the temperature dependence described below.

Figure 2(a) shows the current profile, $I$, observed at $V_{DS} = 300$ μV and $T = 20$ mK, where nonlinear current flows only through the ground state. The saturated current is given by $e\Gamma_L$ in the strong asymmetric barrier condition, while the current profile is given by the thermal distribution of electrons in the electrodes (described later). The derivative of the current, $|dI/dV_L|$, and the integrated current fluctuation $\Delta I$ are also shown in Fig. 2(b) and 2(c), respectively. $\Delta I$ becomes maximum at the edges of the conductive region, and the double-peak structure in the $\Delta I$ curve is very close to the $|dI/dV_L|$ curve. This indicates that the noise is dominated by the fluctuation of the potential ε rather than that of Γ. When the magnitude of the potential fluctuation, Δε, is not very large, we can estimate it by using the simple relation $\Delta I = \alpha^{-1}|dI/dV_L|\Delta\varepsilon$, where $\alpha = 0.14$ eV/V is a conversion factor from the left gate voltage to the potential energy. In this particular condition, we obtained $\Delta\varepsilon = 0.07$ μeV from the peak value at $\mu_d = \varepsilon_{0,0}$ (open circle) and $\Delta\varepsilon = 0.16$ μeV at $\mu_s = \varepsilon_{0,0}$ (solid circle). In addition, small but nonzero current noise is seen in the conductive region between the two peaks in Fig. 2(c). Since $|dI/dV_L|$ is almost zero in this region, this small $\Delta I$ should come from the fluctuation of the tunneling rate, $\Delta\Gamma_L$. We estimated $\Delta\Gamma_L$ to be $2.4 \times 10^5$/s, and then $\Delta\Gamma_L/\Gamma_L \sim 10^{-3}$. Since the tunneling barrier in this device is formed by depleting electrons by applying negative gate voltages, the tunneling rate should also be influenced by the fluctuation of the barrier height. The fluctuation of the dot potential (Δε) may also change the effective barrier height. When the electron traps are uniformly distributed in the device, the fluctuations of the barrier height and dot potential are considered to be the same magnitude and independent. In this case, the observed $\Delta\Gamma_L/\Gamma_L$ and Δε should be related to each other [$(\Delta\Gamma_L/\Gamma_L)/\Delta\varepsilon \sim 10$ (meV)$^{-1}$ from the noise measurement].

This relation can be inferred from two estimations. Tunneling probability, $W$, through a parabolic potential barrier, $U(x) = U_0 - \tfrac{1}{2}m_e\omega_0^2 x^2$, was calculated under the Wentzel-Kramers-Brillouine (WKB) approximation using the realistic parameters ; effective mass $m_e$, the characteristic energy $\hbar\omega_0 = 1 - 3$ meV, and the effective barrier height relative to the Fermi energy $U_0 - E_F = 3 - 5$ meV.[14] The change of tunneling probability, $\Delta W$, for a small change of barrier height $\Delta U_0$ is approximately given by $(\Delta W/W)/\Delta U_0 = 2 - 5$ (meV)$^{-1}$, which is comparable to the observed $(\Delta\Gamma_L/\Gamma_L)/\Delta\varepsilon$. The other estimation was made experimentally from the $V_L$ dependence. A small change of the gate voltage, $\Delta V_L$, induces excess charge on the gate electrode. Since this gate electrode is located almost the same distance from the tunneling barrier (about 100 nm) and from the QD (about 150 nm), the induced charge changes the dot potential and the barrier height by almost the same amount (neglecting the screening effect). From the $V_L$ dependence of the current characteristics, we obtain $(d\Gamma_L/\Gamma_L)/d\varepsilon = 3$ (meV)$^{-1}$, which is also comparable to the noise measurement. This evidence suggests that the fluctuation of tunneling rate originates from the fluctuation of the barrier potential. Therefore, the 1/f current noise can be understood as the fluctuation of the confinement potential around the QD.

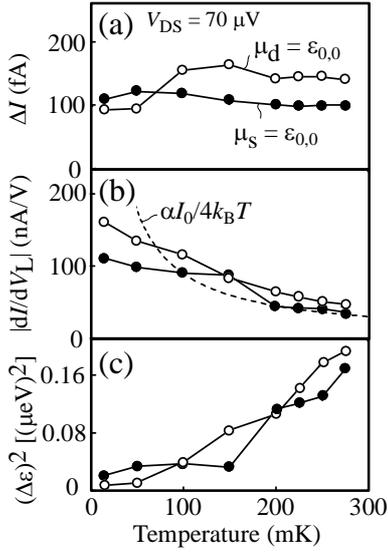

Fig. 3. (a) Temperature dependence of the current fluctuation. Open and solid circles are obtained at $\mu_d = \varepsilon_{0,0}$ and $\mu_s = \varepsilon_{0,0}$, respectively. (b) The derivative of the current spectra. The dashed line is the expected curve from the thermal broadening. (c) Square of the potential fluctuation.

The standard 1/f noise model assumes that electronic traps are uniformly distributed in the device.[7,8] If each trap has its own activation energy and the electron occupation thermally fluctuates, the power spectrum of the charge fluctuation on a trap is a Lorentzian. By assuming that the activation energy is also uniformly distributed in the range of interest, the ensemble power spectrum of charge fluctuation becomes a 1/f spectrum given by

$$S_q \propto \frac{kT}{f}. \qquad (2)$$

This simple model explains the 1/f frequency dependence, and also indicates that the noise power is proportional to the temperature.[12] However, the power of the current noise usually does not follow the linear temperature dependence.[15] We tested how well this model applies to the potential fluctuation in our device.

The temperature dependence of $\Delta I$, $|dI/dV_L|$ and $(\Delta\varepsilon)^2$ is summarized in Fig. 3. A relatively small voltage $V_{DS} = 70$ μV was applied to minimize the influence from the excited states. Data represented by open (solid) circles were obtained at $\mu_d = \varepsilon_{0,0}$ ($\mu_s = \varepsilon_{0,0}$). The current fluctuation $\Delta I$ is almost independent of temperature in the measurement range [See Fig. 3(a)]. However, $\Delta I$ contains the temperature dependence of the device characteristics, which are not included in Eq. 2. The derivative $|dI/dV_L|$ decreases with increasing temperature, as shown in Fig. 3(b). At higher temperature (> 100 mK), this can be explained by the Fermi distribution of electrons in the electrodes. The dashed line shows the expected temperature dependence, $\alpha I_0/4k_BT$, where $I_0 = 23$ pA is the saturation current. At low temperature, $|dI/dV_L|$ may be limited by the lifetime broadening or the effective electron temperature (~ 100 mK). Nevertheless, the current becomes more sensitive to the potential fluctuation at lower temperature. So we evaluated the potential fluctuation in order to disregard the device characteristics. As shown in Fig. 3(c), the power of the estimated potential fluctuation $(\Delta\varepsilon)^2$ has almost linear temperature dependence. Therefore, the simple 1/f noise model is still valid for QD devices at low temperature.

Next, we discuss the influence of the energy relaxation on the background charge fluctuation. When a large $V_{DS}$ is applied, electrons lose their energy somewhere in the device. No influence is expected if the energy relaxation occurs very far from the QD. However, 1/f noise may be enhanced if the energy relaxation takes place near or inside the QD. The energy transfer from conducting electrons to the electron traps has been discussed for a current noise in a metallic SET, but the experimental result was inconclusive.[15] We investigate this effect in the nonlinear transport regime, where energy relaxation process can be controlled by external voltages.

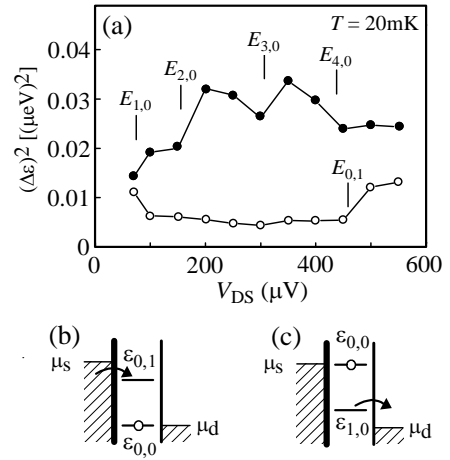

Fig. 4. (a) Source-drain voltage dependence of the potential fluctuation. $E_{i,0}$ and $E_{0,j}$ indicate the threshold voltage for exciting the quantum dot. (b) Energy diagram of single-electron transport at $\mu_d = \varepsilon_{0,0}$. (c) Energy diagram at $\mu_s = \varepsilon_{0,0}$.

We focus on the positive $V_{DS}$ region, 70 μeV < $V_{DS}$ < 600 μeV, where a couple of excited states are involved in the transport. Figure 4(a) shows the $V_{DS}$ dependence of the power of the potential fluctuation $(\Delta\varepsilon)^2$ obtained at $\mu_s = \varepsilon_{0,0}$ and $\mu_d = \varepsilon_{0,0}$. The fluctuation slightly increases with $V_{DS}$, but the change is very small compared to the temperature dependence. At $\mu_d = \varepsilon_{0,0}$, $i$-th excited states of the ($N$+1) electron QD can be occupied at a large excitation voltage, $eV_{DS} > E_{i,0} = \varepsilon_{i,0} - \varepsilon_{0,0}$, as shown by the arrow in the energy diagram of Fig. 4(b). Similarly, $j$-th excited states of the $N$ electron QD can be occupied at $eV_{DS} > E_{0,j} = \varepsilon_{0,0} - \varepsilon_{0,j}$ [See Fig. 4(c)]. The energy spacing $E_{i,0}$ and $E_{0,j}$, obtained from conventional excitation spectra, seem to coincide with the excitation voltages at which the potential fluctuation changes slightly, as indicated by vertical lines in Fig. 4(a). Since no substantial change is observed unless the QD is excited (See open circles for $V_{DS}$ < 400 μV), energy relaxation outside the QD does not cause significant charge fluctuation. In contrast, energy relaxation inside the QD may have influenced the background charge fluctuation. Energy relaxation from the excited state to the ground state accompanies a phonon emission, which might influence the charge distributions near the dot. However, further studies are required in order to clarify the energy transfer mechanism. Since monochromatic phonons are generated locally from a QD by the single-electron tunneling process, interaction with a specific trap may involve some interesting physics.

In practice, the magnitude of the potential fluctuation strongly depends on samples, even though they are fabricated in the same batch. The magnitude also depends on the gate voltages in a sample, indicating a large spatial distribution of trap density. A specific peak with relatively small fluctuations was investigated here. Minimizing trap density will require further study.

In summary, we have investigated 1/f noise in a tunneling current through a GaAs QD. The current noise can be understood as a fluctuation of the confinement potential and tunneling barrier. Our results indicate that the potential fluctuation decreases with decreasing temperature, although the current noise may not be decreased.

*This work was partially supported by BK 21 of Korea.*